\documentclass[conference]{IEEEtran}
\IEEEoverridecommandlockouts
\usepackage{cite}
\usepackage{amsmath,amssymb,amsfonts}
\usepackage{algorithmic}
\usepackage{graphicx}
\usepackage{textcomp}
\usepackage{xcolor}
\usepackage{comment}
\def\BibTeX{{\rm B\kern-.05em{\sc i\kern-.025em b}\kern-.08em
    T\kern-.1667em\lower.7ex\hbox{E}\kern-.125emX}}

\title{Fairness-Aware Network Embeddings: Methods, Applications, and Challenges}

\author{

\IEEEauthorblockN{Ella Has}

\IEEEauthorblockA{\textit{LIACS, Leiden University} \\
Leiden, The Netherlands \\
e.has@liacs.leidenuniv.nl}
\and
\IEEEauthorblockN{Harshith Kumar Yadav}
\IEEEauthorblockA{\textit{Dept. of Computer Science} \\
\textit{IIT Ropar, India}\\
e.23csz0002@iitrpr.ac.in}
\and
\IEEEauthorblockN{Gaurav Dixit}

\IEEEauthorblockA{\textit{Mehta Family School of DS and AI} \\
\textit{IIT Roorkee, India}\\
gaurav.dixit@ms.iitr.ac.in}
\and

\vspace{0.1cm}
\hspace*{2.5cm}
\begin{tabular}[t]{c}
\IEEEauthorblockN{Mykola Pechenizkiy}
\vspace{-0.4cm}
\IEEEauthorblockA{\textit{Eindhoven University of Technology} \\
Eindhoven, The Netherlands \\
m.pechenizkiy@tue.nl}
\end{tabular}

\and

\IEEEauthorblockN{Akrati Saxena}
\IEEEauthorblockA{\textit{LIACS, Leiden University} \\
Leiden, The Netherlands \\
a.saxena@liacs.leidenuniv.nl}
\and
}

\begin{document}

\maketitle

\begin{abstract}
Network embedding methods learn low-dimensional representations of graph-structured data to support downstream tasks such as node classification, link prediction, and influence maximization. However, real-world networks often reflect structural inequalities arising from demographic imbalances, homophily, and other societal biases, which fairness-agnostic embedding methods can encode and amplify. To address this issue, numerous fairness-aware network embedding methods have been proposed to mitigate bias while preserving embedding utility. This survey presents a comprehensive overview of fairness-aware network embeddings for complex networks. We propose a taxonomy that categorizes existing methods along three main complementary dimensions: underlying embedding approach (spectral, random walk, graph neural network, Bayesian, and method-agnostic), fairness intervention strategy (pre-processing, in-processing, and post-processing), and fairness objective criterion (embedding- or task-level). We further compare methods with respect to group versus individual fairness and assumptions regarding sensitive attributes. Finally, we discuss current limitations and highlight promising future research directions. This survey provides a unified perspective on fairness-aware network embedding and serves as a reference for developing fair and trustworthy network representation learning methods. 
\end{abstract}

\section{Introduction}

Networks provide a fundamental abstraction for representing complex systems by modeling entities as nodes and their interactions as edges \cite{newman2018networks}. This representation naturally models a wide range of real-world systems, including social networks, citation networks, biological networks, transportation systems, financial transaction networks, and communication infrastructures. A central objective of network analysis is to extract meaningful information from the network structure to support downstream tasks such as node classification, link prediction, community detection, node ranking, influence maximization, anomaly detection, and graph classification.

\begin{figure}
    \centering
    \includegraphics[width=\linewidth]{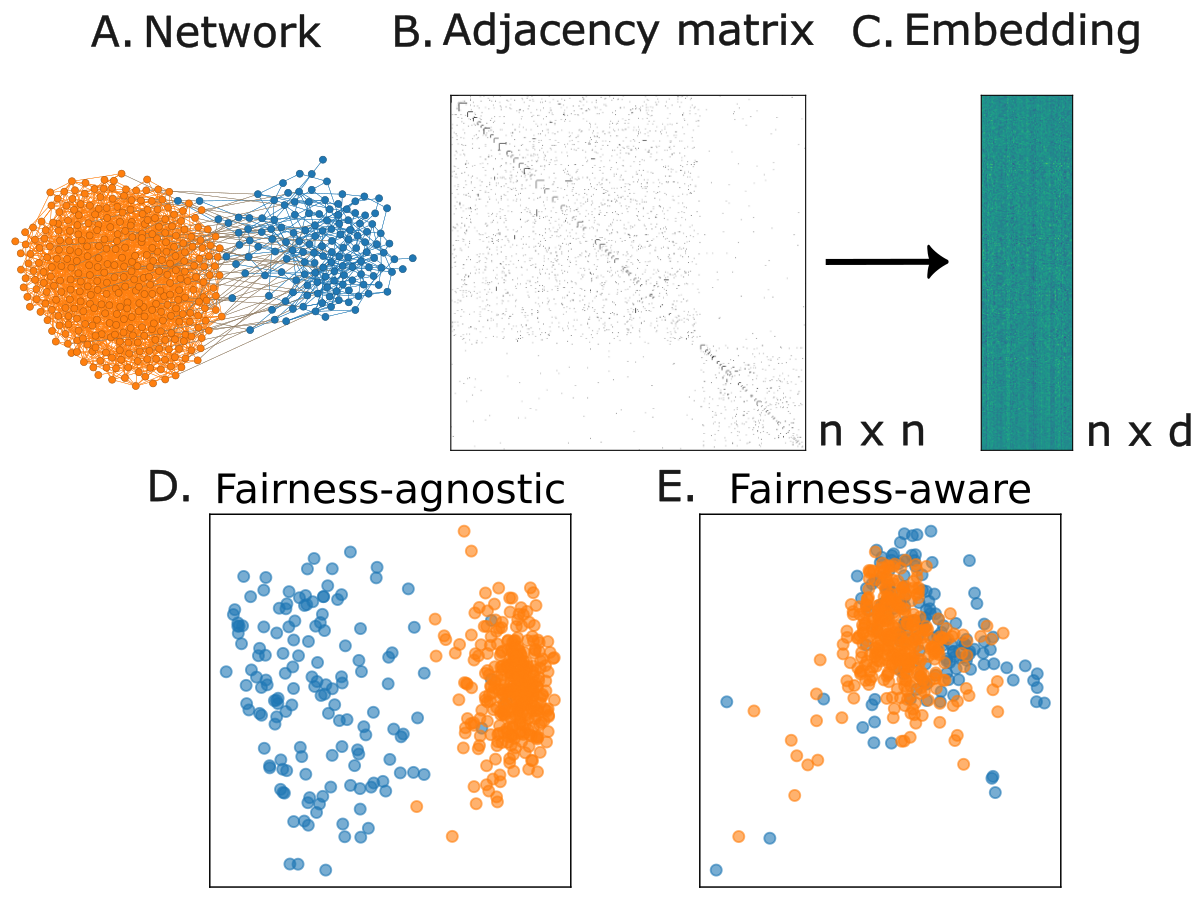}
    \caption{An illustration of fairness-aware network embedding. (A) An example network with two demographic groups (orange and blue). (B) The corresponding adjacency matrix. (C) A sample low-dimensional embeddings. (D) A fairness-agnostic embedding (DeepWalk \cite{perozzi_deepwalk_2014}) that preserves sensitive group information, potentially leading to biased downstream predictions. (E) A fairness-aware embedding (CrossWalk \cite{khajehnejad_crosswalk_2022}) that mitigates the impact of sensitive attribute information, resulting in a fair outcome in downstream tasks.}  
    \label{fig:embeddings}
\end{figure}

Network embedding has emerged as one of the most successful approaches to analyzing graph-structured data \cite{cui2018survey}. The goal of network embedding is to map each node to a low-dimensional vector space while preserving the network's structural and attribute-based information (as shown in Fig.~\ref{fig:embeddings}). These learned representations serve as features for developing machine-learning- and deep-learning-based models for downstream network analysis tasks. Classical embedding techniques include spectral methods \cite{von2007tutorial}, random-walk-based models such as DeepWalk~\cite{perozzi_deepwalk_2014} and Node2vec~\cite{grover_node2vec_2016}, and, more recently, graph neural networks (GNNs), such as GCN~\cite{kipf2016semi} and GraphSAGE~\cite{hamilton2017inductive}, which have become the dominant paradigm for representation learning due to their expressive message-passing mechanism. Node embedding quality directly influences the performance of downstream network analysis tasks. 

Since embeddings are learned from network topology and node attributes, they can also inherit and amplify structural inequalities present in the underlying network, leading to bias in downstream task outcomes (Fig. 1). This may disadvantage minority groups through lower recommendation visibility~\cite{rahman_fairwalk_2019}, lower centrality rankings~\cite{tsioutsiouliklis2021fairness}, reduced information access \cite{saxena2026dq4fairim, saxena2023fairness}, and poorer node classification performance~\cite{dai_say_2021, de2024group}.

Real-world networks exhibit structural inequalities among individuals or groups, shaped by historical, societal, and behavioral processes that influence their evolution \cite{saxena_fairsna_2024, karimi2018homophily, macedo2026gender}. One important source of bias is group-size imbalance, in which minority groups are significantly underrepresented, thereby providing limited information for learning reliable representations~\cite{liu_iagnn_2025}. Another major source is homophily \cite{saxena2025homophily}, the tendency of individuals sharing similar characteristics, such as gender, ethnicity, age, or political affiliation, to connect more frequently with one another. Homophily leads to segregated network structures that reduce interactions between demographic groups and can amplify disparities in downstream predictions~\cite{li_fairlp_2022, karimi2018homophily}. 
Other evolution mechanisms, including preferential attachment \cite{barabasi1999emergence}, reciprocity, community formation, and core-periphery structure \cite{saxena2016evolving}, further reinforce existing inequalities. 

As a result, network topology itself becomes a carrier of sensitive demographic information, leading to undue influence of sensitive attributes on task outcomes. For example, in attributed networks, sensitive attributes are often correlated with both network structure and node features, enabling downstream models to exploit protected attributes even when they are omitted from the input. Therefore, simply removing sensitive labels is insufficient to eliminate bias. These concerns have motivated the rapidly growing field of fairness-aware network embeddings, with the objective of learning representations that maintain predictive utility while reducing disparities across sensitive groups.

In this survey, we present a comprehensive review of fairness-aware network embedding methods for complex networks. We introduce a taxonomy that categorizes existing approaches along three complementary dimensions: the underlying embedding approach, the fairness intervention strategy, and the fairness objective optimized. We further compare methods with respect to group versus individual fairness assumptions about sensitive attributes, and applicability to different network settings. Finally, we discuss the strengths and limitations of existing approaches, identify emerging research trends, and outline important open challenges to guide future research toward fair, scalable, and trustworthy embeddings.

\section{Network Embeddings and Applications} 

Let a graph be represented as $G=(V, E,\mathbf{X})$, where $V$ denotes the set of nodes, $E \subseteq V \times V$ is the set of edges, and $\mathbf{X}$ represents optional node attributes. The objective of network embedding is to learn a mapping
\[
f: V \rightarrow \mathbb{R}^{d},
\]
that assigns each node $v \in V$ a low-dimensional vector representation $\mathbf{z}_v=f(v)$, 
where $d \ll |V|$. The embedding function is learned such that similar nodes in the original graph are mapped to nearby points in the embedding space. Depending on the embedding method, the optimization objective may preserve first-order proximity (direct connections), higher-order neighborhood relationships, structural similarity, community structure, or node attribute information. The learned embedding matrix $\mathbf{Z}\in\mathbb{R}^{|V|\times d}$ provides a compact representation of the graph that can be used as input features for a wide variety of downstream tasks, as explained below.

\paragraph{Node Classification}
Given a set of labeled nodes $\mathcal{L}\subseteq V$, a classifier $g:\mathbb{R}^{d}\rightarrow\mathcal{Y}$ is trained on the corresponding node embeddings to predict labels for unlabeled nodes, where $\mathcal{Y}$ denotes the set of class labels \cite{arya2022node}. Common choices include logistic regression~\cite{lavalley2008logistic, kose_dynamic_2023}, label propagation~\cite{zhu2002learning, khajehnejad_crosswalk_2022}, support vector machines~\cite{hearst1998support}, multilayer perceptrons~\cite{navarin_learning_2020}, and GNNs~\cite{hamilton2017inductive}.

\paragraph{Link Prediction}
The objective of link prediction is to estimate whether an edge exists or is likely to form between two nodes. Given node embeddings, pairwise node features are constructed using operators such as concatenation, Hadamard product, L1 norm, L2 norm, or cosine similarity \cite{grover_node2vec_2016, saxena_nodesim_2022}. These features are then used to train a machine learning model, often logistic regression, that predicts missing or future links.

\paragraph{Node Ranking and Influence Maximization}
Node embeddings capture structural features that can be used to estimate node importance or centrality ranking \cite{saxena2020centrality}. A ranking function $r(v)=h(\mathbf{z}_v)$ assigns an importance score to each node, where $h(\cdot)$ can be a learned regression model. For influence maximization, node embeddings can be clustered using $k$-medoids, and the top-$k$ nodes from each cluster are selected with a group-aware seed selection strategy to maximize the spread of influence \cite{khajehnejad_crosswalk_2022}.

\paragraph{Anomaly Detection}
Network embeddings can also be used to identify anomalous nodes whose structural behavior differs substantially from the majority of the network~\cite{hu2016embedding}. Given the embedding matrix $\mathbf{Z}$, anomaly scores can be computed using clustering, density estimation, nearest-neighbor distances, or reconstruction errors. Nodes with isolated embeddings or high reconstruction errors are identified as potential anomalies.

Network embeddings have also been widely applied to community detection~\cite{sun2020network}, recommendation systems~\cite{ying2018graph}, graph classification~\cite{ma2020graph}, graph matching~\cite{li2019graph}, visualization~\cite{baingana2014embedding}, and network alignment~\cite{chu2019cross}. By providing a low-dimensional representation that captures both local and global structural information, embeddings enable conventional machine learning models to operate effectively on graph-structured data.

\section{Taxonomy}
In this survey, we classify fair network embedding methods along three dimensions: underlying embedding approach, fairness intervention strategy, and fairness objective. Additionally, we discuss group and individual fairness and requirements on the sensitive attributes.

\subsection{Embedding Approach}
Fair network embedding methods can be classified by their underlying embedding approach: spectral embedding, random walk-based embedding, GNNs, Bayesian modeling, and method-agnostic approaches, with GNNs constituting the vast majority. Method-agnostic approaches propose bias-mitigating strategies that can be applied regardless of the training approach. For example, some methods reduce bias by augmenting the input graph~\cite{li_fairlp_2022}, which can then be embedded by any method.

\subsection{Fairness Intervention Strategy}

The second dimension classifies methods according to how and when fairness is introduced into the embedding pipeline, i.e., pre-processing, in-processing, post-processing, and hybrid frameworks~\cite{li_spectral_2023, wu_learning_2021}.

\subsubsection{\textbf{Pre-processing methods}} modify the input graph before learning node embeddings. Their objective is to reduce structural sources of bias, for example, by adding or removing edges to reduce homophily~\cite{cong_fairsample_2024, spinelli_fairdrop_2022} or rebalance connectivity between demographic groups~\cite{li_fairlp_2022}.

\subsubsection{\textbf{In-processing methods}} incorporate fairness directly into the embedding process. Existing approaches achieve this through three main mechanisms: graph augmentation, rebalancing, and objective function modification. Graph augmentation dynamically perturbs the graph during training, generating different graph views to encourage fair representations~\cite{agarwal_towards_2021}. Rebalancing techniques adjust the learning process by modifying sampling probabilities~\cite{cao_flexible_2023}, random walk transitions~\cite{rahman_fairwalk_2019, khajehnejad_crosswalk_2022}, or GNN message passing~\cite{zhu_fairagg_2024, lin_bemap_2024} and attention~\cite{kose_fairgat_2024, luo_fairgt_2024} weights to ensure a more balanced representation of different groups. 
Fairness can be incorporated into the objective function through:
\renewcommand{\labelenumi}{\alph{enumi})}
\begin{enumerate}
    \item \textbf{Constraints:} restricting the search space~\cite{kleindessner_guarantees_2019};
    \item \textbf{Regularization:} penalizing bias metrics~\cite{kose_dynamic_2023, agrawal_no_2024};
    \item \textbf{Adversarial learning:} obscuring sensitive attributes from an adversarial discriminator~\cite{bose_compositional_2019, dai_say_2021};
    \item \textbf{Contrastive learning:} aligning embeddings across fairness-aware graph views~\cite{kose_fair_2022};
    \item \textbf{Disentanglement:} separating task-relevant and sensitive information~\cite{guo_towards_2023, zhu_fair_2024}.
\end{enumerate}

\subsubsection{\textbf{Post-processing methods}} improve fairness after training. They either transform the learned embeddings (e.g., by decorrelating them from sensitive attributes~\cite{palowitch_debiasing_2020}) or adjust the outputs of downstream tasks to produce fair decisions without retraining the model~\cite{li_spectral_2023, fan_fair_2025}.

\subsection{Fairness Objective} 

The third dimension of our taxonomy categorizes methods according to the fairness objective they optimize. Existing methods generally target fairness at either (i) the embedding level or (ii) the downstream task level.

\subsubsection{\textbf{Embedding-level fairness}} considers to what extent embeddings encode sensitive attributes. This is typically evaluated with the correlation between node embeddings and sensitive attributes or the Representation Bias (RB), which quantifies the ability of a classifier to predict sensitive attributes from the embeddings. Lower correlation or prediction accuracy indicates that the embeddings encode less sensitive information.

\subsubsection{\textbf{Task-level fairness}} evaluates whether the downstream predictions are fair across demographic groups. Common fairness metrics include statistical parity (SP), equality of opportunity (EO), and performance disparity across groups. SP requires favorable outcomes to be independent of sensitive attributes, while EO requires equal true positive rates across groups. Performance disparity measures differences in predictive performance across demographic groups, for example, using variance. Additionally, several methods optimize task-specific fairness objectives that may not generalize to fairness in other tasks, such as balanced cluster composition~\cite{kleindessner_guarantees_2019, gupta_consistency_2022} or increased inter-group link prediction~\cite{rahman_fairwalk_2019, saxena_nodesim_2022}.

\subsection{Additional Methodological Characteristics} 

Beyond the three primary dimensions of our taxonomy, fairness-aware network embedding methods also differ in several methodological characteristics, as discussed below. 
    \subsubsection{\textbf{Group vs. individual fairness}} Most approaches aim to reduce disparities between demographic groups (group fairness). In contrast, a smaller number of methods focus on individual fairness~\cite{gupta_consistency_2022, dong_individual_2021}, ensuring that structurally or semantically similar nodes receive similar outcomes.
    \subsubsection{\textbf{Binary vs. multi-class sensitive attributes}} Many existing methods assume a single binary sensitive attribute for simplicity. However, some approaches naturally support multi-class sensitive attributes~\cite{moens_re_2023} or can be extended to settings involving more than two demographic groups.
    \subsubsection{\textbf{Single vs. multiple sensitive attributes}} While most methods consider only one sensitive attribute, some address \textit{multiple} or \textit{intersectional} attributes (e.g., gender and ethnicity simultaneously), enabling fairness across overlapping demographic groups~\cite{buyl_debayes_2020, wu_learning_2021}.
    \subsubsection{\textbf{Availability of sensitive attributes}} Most fairness-aware embedding methods assume that sensitive attributes are available for all nodes during training. More recent approaches relax this assumption by handling partially observed sensitive attributes~\cite{dai_say_2021, guo_fair_2023} or eliminating the need for them altogether through feature-blind learning~\cite{wang_towards_2025}.

The upcoming sections (\ref{spectral}-\ref{agnostic}) cover fair embedding methods categorized by their underlying approach.

\section{Spectral Embedding}\label{spectral}

Spectral embedding methods learn node representations by performing eigenvalue decomposition on the network's Laplacian matrix and using the eigenvectors corresponding to its smallest eigenvalues as embeddings~\cite{von2007tutorial}. Fairness-aware spectral embedding methods primarily improve fairness through fairness constraints or regularization.

\subsection{Fairness Constraints}

Fairness constraints in spectral embedding encourage balanced node representations across sensitive groups. Kleindessner et al.~\cite{kleindessner_guarantees_2019} formulated fair spectral clustering by constraining the optimization so that each cluster reflects the demographic composition of the network. This is achieved by transforming the graph Laplacian using the proportional group sizes before eigenvalue decomposition. Wang et al.~\cite{wang_scalable_2023} later improved the scalability of this approach. Beyond group fairness, Gupta and Dukkipati~\cite{gupta_consistency_2022} extended spectral clustering to individual fairness by constraining nodes to receive similar representation in each cluster. FNM~\cite{li_spectral_2023} further combines fairness-constrained spectral embedding with balanced $k$-means clustering, enforcing demographic balance throughout the clustering process.

\subsection{Fairness Regularization}

DFaR~\cite{kose_dynamic_2023} adopts a regularization approach that penalizes the correlation between node embeddings and sensitive attributes. It first learns node and feature embeddings using unconstrained spectral embedding, and then applies a correlation penalty to learn the optimal merging of node and feature embeddings. DFaR can also be used for dynamic networks.

\section{Random Walk-based Embedding}

Random walk-based embedding methods, also known as skip-gram embedding, such as DeepWalk~\cite{perozzi_deepwalk_2014} and Node2vec~\cite{grover_node2vec_2016}, generate random walks to capture the context of each node and learn embeddings using the skip-gram model~\cite{mikolov2013efficient}, which distinguishes context nodes from randomly sampled negative nodes. Fairness-aware variants primarily mitigate bias by modifying either the random-walk sampling process or the negative-sampling strategy.

\subsection{Rebalancing}
 
Most fairness-aware random walk methods improve fairness by rebalancing random walk sampling to collect an equitable context for each node. Fairwalk~\cite{rahman_fairwalk_2019} balances random walk transition probabilities across all groups, ensuring equal group representation in sampled walks. However, Fairwalk does not account for the possibility that nodes might have neighbors only of the same group, causing walks to remain stuck within one group and over-represent it. CrossWalk~\cite{khajehnejad_crosswalk_2022} addresses this limitation by biasing walks toward nodes with more diverse neighborhoods near group boundaries, while MoonWalk~\cite{moens_re_2023} extends this idea to multi-group settings. NodeSim~\cite{saxena_nodesim_2022} introduces a similarity-aware random walk based on node similarity and community structure to better capture inter-community relationships, with fairness evaluated through inter- and intra-community link prediction.

\subsection{Negative Sampling}
Unlike sampling-based approaches, residual2vec~\cite{kojaku_residual2vec_2021} mitigates bias by modifying the negative sampling distribution. It generates randomized graphs that preserve only bias-inducing properties, such as node degree and inter-/intra-group connectivity, and draws negative samples according to node-pair frequencies in these graphs, filtering out the effect of sensitive-attribute homophily on random walks by sampling nodes from the same group more often as negative pairs while providing a flexible framework that could model any bias-inducing properties in the randomized graphs. 

\section{Graph Neural Networks}

Graph neural networks (GNNs) are deep learning models specifically designed for graph-structured data that learn embeddings by jointly exploiting node features and graph topology~\cite{kipf2016semi, hamilton2017inductive}. At each layer, a node aggregates its neighbors' features, known as message passing. The parameters of the aggregation function are trained for optimal embedding utility. After multiple layers, the node embeddings capture both local neighborhood information and higher-order structural patterns, making them highly effective for downstream tasks. 

\subsection{Rebalanced Message Passing}
Fair message passing is an in-processing approach that mitigates bias during neighborhood aggregation, where graph structure can propagate and amplify sensitive information. 

\subsubsection{\textbf{Fair Message Aggregation}}
Fair message aggregation approaches rebalance the contribution of each sensitive group, either by sampling an equal number of neighbors per group~\cite{lin_bemap_2024} or by reweighting each group's contribution. FairAGG~\cite{zhu_fairagg_2024} uses Shapley values to estimate the fairness contributions of intra- and inter-group edges and reweights their messages accordingly. FAME~\cite{purificato_gnns_2025} offers a lightweight solution that directly modifies each GCN message or GAT attention logit in A-FAME according to the difference in sensitive attributes between connected nodes. DegFairGNN~\cite{liu_generalized_2023} improves degree fairness by learning contexts that enrich information for low-degree nodes and distill information for high-degree nodes. Fair Graph U-net~\cite{wang_fair_2025} introduces fairness into hierarchical representation learning by balancing group representation during pooling, while distributing the cost of group fairness more evenly across individuals. Im-GBK~\cite{liang2023heterophily} handles homophilic and heterophilic neighbors separately.

HetroFair~\cite{gholinejad_heterophily-aware_2026} reduces popularity bias in a recommendation system by inversely weighting user–item messages based on their similarity, suppressing popular items and strengthening long-tail contributions. IntFair~\cite{guo_intfairgraph_2025} promotes consistent recommendation performance across categories preferred by users through TOPSIS-based~\cite{hwang2012multiple} balanced neighbor sampling. 

\subsubsection{\textbf{Fair Attention Weights}}
FairGAT~\cite{kose_fairgat_2024} allocates theoretically determined attention budget for inter-group neighbors. FairGT~\cite{luo_fairgt_2024} extends this by constructing sensitive-aware multi-hop token sequences and combines them with selected adjacency eigenvectors before Transformer attention.

\subsection{Fairness Regularization}
Another common in-processing approach in fair GNNs is fairness regularization, which adds some measurement of bias as a penalty to the loss function. Unfairness is penalized by measuring bias at the embedding or at the outcome level.

\subsubsection{\textbf{Embedding-Level Penalties}}
DFGNN~\cite{navarin_learning_2020} proposes two ways to measure unfairness as the distance between embedding distributions between groups: comparing a summary metric by taking the difference in means, later also used by FairNorm~\cite{kose_fairnorm_2022}, or comparing the shape of the distributions with Wasserstein distance~\cite{kantorovich1939mathematical}. 
FairGAE~\cite{fan_fair_2021} pairs Wasserstein distance penalization with balanced message passing. FairHGNN~\cite{cao_flexible_2023} uses it in a GNN for Heterogeneous Information Networks by sampling meta-paths with balanced probabilities for each group. FAHIN~\cite{cao_fahin_2024} additionally adaptively changes the importance of the fairness regularization term.

Others have suggested different metrics. GMMD~\cite{zhu_fairness-aware_2023} uses the summary metric Maximum Mean Discrepancy (MMD)~\cite{NEURIPS2019_944a5ae3}. FairGLite~\cite{wang_fairness-aware_2026} learns a mask that hides sensitive attributes with one penalty term penalizing the distance between embedding distributions of groups, and another penalizing correlation with sensitive labels. Missing labels are first estimated, and nodes with more certain labels are more heavily regularized. EAGNN~\cite{zhang_towards_2026} compares fairness at both the embedding and the task level with regularization terms minimizing embedding distance between similar nodes from different groups and maximizing statistical parity. In contrast, the feature-blind method Fairwos~\cite{wang_towards_2025-1} penalizes misalignment between embeddings of similar nodes by extracting pseudo-sensitive attributes from graph structure and node features.

\subsubsection{\textbf{Task-Level Penalties}}
On the other hand, FMP~\cite{jiang_chasing_2024} penalizes disparities at the task outcome level using statistical parity (SP), while FS-GNN~\cite{zhao2025fs} penalizes the difference in utility loss between groups. Agrawal et al.~\cite{agrawal_no_2024} applied equality of opportunity (EO) as their measure of fairness in a federated learning setting where embeddings are updated locally to preserve privacy, while EO is estimated in a privacy-preserving way by aggregating noisy performance and group membership data across users. REDRESS~\cite{dong_individual_2021} ensures individual fairness by penalizing the distance between the ranking of nodes with the most similar characteristics and the ranking of nodes with the most similar outcomes. ComFairGNN~\cite{sium_comfairgnn_2025} debiases by encouraging nodes with the same class label across groups and local structures to receive comparable embeddings regardless of sensitive attributes. In the case of unavailable sensitive labels, FairINV~\cite{zhu_one_2024} identifies group partitions where the GNN exhibits performance disparities and penalizes loss variation across these sensitive group partitions.

\subsection{Adversarial Learning}

Adversarial fair GNNs learn node embeddings that remain informative for downstream tasks while preventing sensitive attribute leakage by obscuring the sensitive information from an adversarial model attempting to decode it. Bose et al.~\cite{bose_compositional_2019} adversarially trained attribute-specific MLP filters to remove sensitive attributes from embeddings. Similarly, Khajehnejad et al.~\cite{khajehnejad_adversarial_2021} used an adversarial autoencoder. FairVGNN~\cite{wang_improving_2022} addresses sensitive attribute leakage by adversarially learning multiple masked feature views and adaptively weight-clamping to limit sensitive-related feature channels.

FairGNN~\cite{dai_say_2021} extends adversarial learning to graphs with limited available sensitive labels by estimating missing sensitive attributes and combining adversarial debiasing with covariance regularization, while FairAC~\cite{guo_fair_2023} combines attention-based attribute completion with adversarial learning. NT-FairGNN~\cite{dai_learning_2023} extends FairGNN by privatizing sensitive attributes and estimates them using a noise-corrected loss. FairHELP~\cite{cao_fairhelp_2023} adapts adversarial learning for Heterogeneous Information Networks by making the adversary predict the sensitive attributes involved in the edges.

FPGNN~\cite{zhang_fpgnn_2023} combines adversarial representation learning with a novel Fair Path pruning strategy to mitigate sensitive information leakage from high-degree nodes that are identified through fair random walks. MVFGNN~\cite{zhang_multi-view_2024} combines original, diffusion and feature-similarity graph views through a variational graph autoencoder (VGAE) and similarity-based weighting before adversarially removing sensitive information.

\subsection{Contrastive Learning}

Fair GNNs based on contrastive learning learn bias-resistant node embeddings by aligning representations across fairness-aware graph views while separating unrelated nodes. Kose et al.~\cite{kose_fair_2022} created graph views with edge deletion and feature masking to reduce biased connectivity patterns. FairMigration~\cite{hu_migrate_2024} pretrains a GNN on counterfactual views with flipped sensitive attributes and adversarially removes residual sensitive information from these migrated groups. FairMIB~\cite{liu_learning_2025} addresses multiple sources of bias by separating the graph into a feature, structural, and diffusion view. Independent variational encoders learn these views, which are then aligned with cross-view contrastive learning. FairDGE~\cite{li_toward_2024} maintains degree fairness in dynamic graphs, where node degrees change over time, using contrastive learning and group-wise utility loss alignment. FairGCL~\cite{dam_fairgcl_2026} applies the contrastive learning framework to influence maximization using a Principal Neighbourhood Aggregation (PNA) encoder and a task-specific loss.





\subsection{Disentanglement}

Disentanglement-based GNNs separate task-relevant information from sensitive information within node representations. CAF~\cite{guo_towards_2023} does so by selecting counterfactual proxy nodes with comparable content but distinct sensitive environments to estimate which components of the embeddings relate to sensitive attributes. SCCAF~\cite{kejani_fair_2024} extends CAF to enhance the discriminative quality of disentangled embeddings. FDGNN~\cite{wang_advancing_2024} considers sensitive information from both node attributes and neighborhoods by finding counterfactual ego graphs in the network to disentangle sensitive and non-sensitive latent factors.

Instead of counterfactual samples, FairSAD~\cite{zhu_fair_2024} learns masks to decorrelate independent latent features and sensitive attributes. FairGID~\cite{chen_learning_2026} also masks sensitive attributes in the features and then learns a structural representation separately. The two representations are then combined with adversarial learning, further reducing sensitive attribute leakage. Similarly, DAB-GNN~\cite{lee_disentangling_2025} disentangles attribute, structural, and attribute-structure interaction biases independently, while Wasserstein distance is used to align embedding distributions across groups. FairMI~\cite{zhao_fair_2023} disentangles embeddings by training a discriminator with the sensitive component and minimizing mutual information with the sensitive-free component.

FGLISA~\cite{wang_fair_2026} addresses unknown sensitive attributes with a causal variational graph encoder that separates sensitive-related and sensitive-unrelated representations, infers soft sensitive labels, and aligns group distributions. FairGNN-WOD~\cite{wang_fairgnn-wod_2025} needs no sensitive attribute labels at all with a VAE that infers sensitive attribute proxies, while Themis~\cite{wang_towards_2025} estimates reliable proxies with a causal Bayesian VAE that separates sensitive and task-related information. 

\subsection{Graph Augmentation}
Graph augmentation in GNNs can be used as a pre- or in-processing step to reduce bias in the input graph or to gain robustness against perturbation of sensitive attributes.

\subsubsection{\textbf{Reduced Bias in Input Graph}}
Several approaches propose GNN pipelines that perturb the input graph to reduce bias. EDITS~\cite{dong_edits_2022} perturbs both the adjacency matrix and the node features to minimize the distance between the feature distributions of two sensitive groups and the distance between feature distributions after rounds of message-passing.

G-FAME~\cite{liu_fair_2023} pre-processes the graph with edge deletion to reduce homophily and addresses the loss of information inherent in removing edges from the graph with a mixture of experts more capable of learning from this limited information. FairSample~\cite{cong_fairsample_2024} connects nearby nodes with similar features and the same (known or estimated) class label but different sensitive attributes and then performs balanced message-passing learned with reinforcement learning. Geb\cite{wang_generating_2024} uses evolutionary search to find unfair subgraphs in a trained GNN. It modifies edges and node features to reduce group differences and updates only the impacted nodes. 

\subsubsection{\textbf{Improved Robustness}}
Another use of graph augmentation is improving robustness against perturbation of the sensitive attributes. Nifty~\cite{agarwal_towards_2021} generates two alternative graphs at each training step, one with augmented edges and node features for learning stability and one with augmented sensitive attributes for fairness. A GNN is trained to achieve similarity between the embeddings of the original and the two augmented graphs. Similarly, AdaLipGNN~\cite{singh_unified_2025} achieves fairness as a byproduct of robustness to graph and feature perturbations. FairCNCB~\cite{xiao_towards_2025} generates realistic counterfactual nodes by changing sensitive attributes while preserving labels and graph structure. It prioritizes minority groups during training and ensures consistent predictions for original and generated nodes. Ma et al.~\cite{ma_learning_2022} also included the influence of the neighbors' sensitive attributes on a node’s embeddings with a causal model that generates counterfactual ego graphs in which either the sensitive attributes of neighbors or of the central node are changed, and the node features are altered accordingly.

\subsection{Hybrid Methods}

Hybrid approaches combine the methods discussed above. RFCGNN+~\cite{wang_toward_2024} combines multi-frequency message passing to reduce topology bias and counterfactual nodes to disentangle sensitive information, while reducing differences between sensitive groups. FDGNN~\cite{zhang_disentangled_2025} uses counterfactual augmentation and disentangled contrastive learning to separate task-relevant representations from sensitive components. MAPPING~\cite{song_mapping_2024} debiases both node features and graph topology before training with fairness regularization, adversarial feature reconstruction, fair message passing, and edge pruning. SRGNN\cite{zhang_learning_2024} augments the graph by strengthening low-degree nodes and reducing high-degree connections and uses an adversarial discriminator to remove sensitive information from embeddings.

\subsection{Other Methods}
The following methods propose unique strategies that do not fit into the categories above. 
FairDTD~\cite{li_toward_2026} reduces feature and topology bias by transferring fairness knowledge from MLP and GCN teacher models trained on partial data into a student GNN. FairGKD~\cite{zhu_devil_2024} combines the knowledge of both teachers into a single synthetic teacher. FairGE~\cite{luo_fairge_2026} pre-processes the data by zero-padding missing sensitive attributes and filtering out sensitive information with spectral truncation of structural encoding as input to a Graph Transformer.

\section{Bayesian Modeling}
Conditional Network Embedding~\cite{kang2018conditional} learns node embeddings using Bayes' rule.
DeBayes~\cite{buyl_debayes_2020} introduces a fair prior that encodes both structural properties (e.g., degree) and sensitive attributes so the likelihood does not need to represent the sensitive attributes for a good fit with the network. The learned likelihood and a prior with only structural properties define the embeddings, filtering out all sensitive information. 

\section{Method-Agnostic Approaches}\label{agnostic}
Unlike approaches that modify a specific embedding method, method-agnostic approaches introduce fairness strategies with no requirements on the model's architecture.

\subsection{Graph Augmentation}
A common method-agnostic bias mitigation strategy is to alter the input graph so that the effects of sensitive attributes on the graph structure are reduced. FairLP~\cite{li_fairlp_2022} pre-processes the input graph by adding and removing edges to achieve equal density in each group because this is an important factor in link prediction accuracy disparity. FairDrop~\cite{spinelli_fairdrop_2022} takes an in-processing approach, removing edges between nodes in the same group from the original graph at each training step to reduce homophily. This can be applied to any embedding backbone that learns in multiple iterations, including random-walk methods and GNNs. Instead of augmenting graphs using heuristics such as reducing homophily, Ling et al.~\cite{ling_learning_2022} trained a GNN model, Graphair, with adversarial learning to produce augmented graphs that hide sensitive attributes while keeping similar network structure, node features, and final embeddings.

\subsection{Filtering Sensitive Attributes}
Another approach is to filter out sensitive information from the embeddings. MONET~\cite{palowitch_debiasing_2020} is an in-processing approach that orthogonalizes the embeddings of sensitive attributes and topological embeddings with eigenvalue decomposition at every training step. On the other hand, Kose et al.~\cite{kose_fairness-aware_2023} designed a graph filter approach that can be applied as a pre- or post-processing step. A graph signal (e.g., input node attributes or output embeddings or labels) is transformed into the frequency domain to filter out part of the signal that corresponds to the sensitive attributes. The post-processing approach FairGo~\cite{wu_learning_2021} learns a filter for each dimension of the sensitive attributes with adversarial learning to obscure sensitive information.

\subsection{Fairness Regularization}
FairMILE~\cite{he_fairmile_2023} uses a combination of pre- and post-processing by first merging strongly connected nodes with different sensitive attributes into diverse supernodes and then learning embeddings on the coarsened graph. The coarse embeddings are refined to obtain embeddings for each node in the original graph with an objective function that penalizes the distance between embeddings in different groups.

\subsection{Task-Level Debiasing}
Fairness constraints can also be enforced at the task level.  FSGNN~\cite{fan_fair_2025} applies this idea to influence maximization by selecting nodes from each group proportionally to the estimated total influence of that group.

\section{Emerging Trends and Future Directions} 

In this section, we discuss emerging trends and suggestions for future research.

\subsection{Sensitive Attributes: Missingness and Intersectionality}
Existing methods largely rely on sensitive attributes during training, which are often unavailable due to privacy or data limitations. Developing feature-blind methods that detect and mitigate structural bias remains a key challenge. Methods that address (partial) missing sensitive attributes usually do so by predicting them before applying fairness constraints~\cite{dai_say_2021, wang_fairgnn-wod_2025}. However, inaccurate predictions may introduce additional bias, particularly when missingness is not random and disproportionately affects minority groups. Future methods should explicitly model uncertainty and missingness mechanisms rather than relying solely on imputed sensitive labels. 

Additionally, most existing methods consider a single binary sensitive attribute, whereas real-world applications involve multiple interacting demographic characteristics. Extending fairness objectives to intersectional and multi-class settings remains a significant challenge~\cite{martin2025intersectional, buyl_debayes_2020, wu_learning_2021}. 

\subsection{Understanding Bias}
Many existing approaches mitigate bias without explicitly modeling how it arises within the network~\cite{dai_say_2021, navarin_learning_2020}. Conversely, graph augmentation and rebalancing methods often rely on assumptions regarding homophily, structural imbalance, or preferential attachment~\cite{li_fairlp_2022, spinelli_fairdrop_2022}. Understanding whether fairness interventions should explicitly model bias-generating mechanisms or remain agnostic to them is an important theoretical question that deserves further investigation. 

Additionally, developing interpretable methods that explain fair decisions will offer insights into the impact of structural biases and the effects of fairness interventions, and help design fair explainable methods. Relatedly, a better understanding of the fairness-utility trade-off remains a key challenge. Developing principled optimization methods and theoretical guarantees for navigating this trade-off rather than relying on hyperparameter optimization will benefit the field.

\subsection{Novel Strategies for Fair Random Walk-based Embeddings} 
Existing fair random walk-based methods modify transition probabilities or sampling strategies. Extending techniques such as fairness regularization and adversarial learning, which have proven effective in GNNs, represents a promising direction for improving fairness in random walk-based embeddings.

\subsection{Scalability, Higher-order Networks and Generalization}
Existing methods usually scale poorly to large networks and are only applicable to static homogeneous networks with pairwise interactions. Additional optimization objectives and fairness constraints often limit scalability. Developing scalable fair algorithms while maintaining computational efficiency remains an important research direction~\cite{kose_fairnorm_2022, wang_scalable_2023}. Additionally, real-world networks evolve over time, yet fair embedding methods that efficiently adapt to dynamic networks remain relatively underexplored~\cite{kose_dynamic_2023, li_toward_2024}. Moreover, extending embedding methods to heterogeneous networks with multiple node and edge types~\cite{cao_fairhelp_2023, cao_flexible_2023} as well as hypergraphs and other higher-order network structures, remains largely unexplored. Furthermore, most existing GNNs and spectral methods optimize fairness for a specific downstream task, limiting their generalizability. Developing fair embeddings that preserve fairness across diverse downstream network analysis tasks would improve practical applicability.

\subsection{Benchmark Datasets and Standardized Evaluation}
Existing methods are evaluated on different datasets, tasks, and fairness metrics, hindering meaningful comparisons. The community would benefit from standardized benchmark datasets, evaluation protocols, and reproducible experimental settings specifically designed for fair embeddings~\cite{qian_addressing_2024}.

\section{Conclusion}

Network embeddings have become a fundamental building block for analyzing graph-structured data and performing downstream network analysis tasks, making fairness an increasingly important consideration in their development. In this survey, we presented a comprehensive overview of fairness-aware network embedding methods and organized the literature through a taxonomy based on the underlying embedding approach, fairness intervention strategy, and fairness criterion. We further compared existing methods with respect to their fairness assumptions, reliance on sensitive attributes, and applicability to different downstream tasks.

Our review highlights that although significant progress has been made, current methods remain concentrated on static, homogeneous networks and predominantly assume complete knowledge of binary sensitive attributes. Moreover, many approaches face challenges in balancing fairness, utility, scalability, and generalizability across downstream tasks. We discuss emerging trends and future directions, which will guide the development of the next generation of fair, robust, and trustworthy network embedding methods. 

\clearpage

\bibliographystyle{ieeetr}
\bibliography{bibliography}

@inproceedings{rahman_fairwalk_2019,
	title = {Fairwalk: Towards Fair Graph Embedding},
    booktitle={Proceedings of the 28th International Joint Conference on Artificial Intelligence},
	eventtitle = {International Joint Conference on Artificial Intelligence ({IJCAI})},
	publisher = {{CISPA}},
	author = {Rahman, Tahleen and Surma, Bartlomiej and Backes, Michael and Zhang, Yang},
	urldate = {2025-11-18},
	year = {2019},
}

@inproceedings{khajehnejad_crosswalk_2022,
	title = {{CrossWalk}: Fairness-Enhanced Node Representation Learning},
	volume = {36},
	rights = {Copyright (c) 2022 Association for the Advancement of Artificial Intelligence},
	issn = {2374-3468},
	url = {https://ojs.aaai.org/index.php/AAAI/article/view/21454},
	doi = {10.1609/aaai.v36i11.21454},
	shorttitle = {{CrossWalk}},
	pages = {11963--11970},
	number = {11},
	booktitle = {Proceedings of the {AAAI} Conference on Artificial Intelligence},
	author = {Khajehnejad, Ahmad and Khajehnejad, Moein and Babaei, Mahmoudreza and Gummadi, Krishna P. and Weller, Adrian and Mirzasoleiman, Baharan},
	urldate = {2025-11-18},
	year = {2022},
	langid = {english},
}

@inproceedings{moens_re_2023,
	title = {[Re] {CrossWalk} Fairness-enhanced Node Representation Learning},
	rights = {Creative Commons Attribution 4.0 International, Open Access},
	url = {https://zenodo.org/record/8173747},
	doi = {10.5281/ZENODO.8173747},
    booktitle={ML Reproducibility Challenge 2022},
	eventtitle = {{ML} Reproducibility Challenge 2022},
	publisher = {Zenodo},
	author = {Moens, Gijs Joppe and De Witte, Job and Göbel, Tobias Pieter and Van Den Oever, Meggie},
	editor = {Sinha, Koustuv and Bleeker, Maurits and Bhargav, Samarth},
	urldate = {2025-11-20},
	year = {2023},
	langid = {english},
}

@inproceedings{wang_towards_2025,
	title = {Towards Fair Graph Learning without Demographic Information},
    booktitle={The 28th International Conference on Artificial Intelligence and Statistics},
	volume = {258},
	eventtitle = {The 28th International Conference on Artificial Intelligence and Statistics},
	pages = {2107--2115},
	publisher = {{PMLR}},
	author = {Wang, Zichong and Hoang, Nhat and Zhang, Xingyu and Bello, Kevin and Zhang, Xiangliang and Iyengar, Sundararaja Sitharama and Zhang, Wenbin},
	year = {2025},
}

@inproceedings{kose_dynamic_2023,
	title = {Dynamic Fair Node Representation Learning},
	issn = {2379-190X},
	url = {https://ieeexplore.ieee.org/abstract/document/10094834},
	doi = {10.1109/ICASSP49357.2023.10094834},
	eventtitle = {{ICASSP} 2023 - 2023 {IEEE} International Conference on Acoustics, Speech and Signal Processing ({ICASSP})},
	pages = {1--5},
	booktitle = {{ICASSP} 2023 - 2023 {IEEE} International Conference on Acoustics, Speech and Signal Processing ({ICASSP})},
	publisher = {{IEEE}},
	author = {Kose, Oyku Deniz and Shen, Yanning},
	urldate = {2026-03-10},
	year = {2023},
	note = {{ISSN}: 2379-190X},
}

@inproceedings{liang2023heterophily,
  title={Heterophily-Based Graph Neural Network for Imbalanced Classification},
  author={Liang, Zirui and Li, Yuntao and Huang, Tianjin and Saxena, Akrati and Pei, Yulong and Pechenizkiy, Mykola},
  booktitle={International Conference on Complex Networks and Their Applications},
  pages={74--86},
  year={2023},
  organization={Springer}
}

@article{martin2025intersectional,
  title={Intersectional inequalities in social ties},
  author={Martin-Gutierrez, Samuel and Cartier van Dissel, Mauritz N and Karimi, Fariba},
  journal={Science Advances},
  volume={11},
  number={45},
  pages={eadu9025},
  year={2025},
  publisher={American Association for the Advancement of Science}
}

@article{macedo2026gender,
  title={Gender biases in online communication: A case study of soccer},
  author={Macedo, Mariana and Saxena, Akrati},
  journal={Applied Intelligence},
  volume={56},
  number={1},
  pages={33},
  year={2026},
  publisher={Springer}
}

@article{saxena2023fairness,
  title={Fairness-aware fake news mitigation using counter information propagation},
  author={Saxena, Akrati and Guti{\'e}rrez Bierbooms, Cristina and Pechenizkiy, Mykola},
  journal={Applied Intelligence},
  volume={53},
  number={22},
  pages={27483--27504},
  year={2023},
  publisher={Springer}
}

@article{saxena2020centrality,
  title={Centrality measures in complex networks: A survey},
  author={Saxena, Akrati and Iyengar, Sudarshan},
  journal={arXiv preprint arXiv:2011.07190},
  year={2020}
}

@incollection{arya2022node,
  title={Node classification using deep learning in social networks},
  author={Arya, Aikta and Pandey, Pradumn Kumar and Saxena, Akrati},
  booktitle={Deep learning for social media data analytics},
  pages={3--26},
  year={2022},
  publisher={Springer}
}

@article{barabasi1999emergence,
  title={Emergence of scaling in random networks},
  author={Barab{\'a}si, Albert-L{\'a}szl{\'o} and Albert, R{\'e}ka},
  journal={science},
  volume={286},
  number={5439},
  pages={509--512},
  year={1999},
  publisher={American Association for the Advancement of Science}
}

@inproceedings{saxena2016evolving,
  title={Evolving models for meso-scale structures},
  author={Saxena, Akrati and Iyengar, SRS},
  booktitle={2016 8th international conference on communication systems and networks (COMSNETS)},
  pages={1--8},
  year={2016},
  organization={IEEE}
}

@inproceedings{de2024group,
  title={Group fairness metrics for community detection methods in social networks},
  author={De Vink, Elze and Saxena, Akrati},
  booktitle={International Conference on Complex Networks and Their Applications},
  pages={43--56},
  year={2024},
  organization={Springer}
}

@article{saxena2026dq4fairim,
  title={DQ4FairIM: Fairness-Aware Influence Maximization Using Deep Reinforcement Learning},
  author={Saxena, Akrati and Yadav, Harshith Kumar and Rutten, Bart and Jha, Shashi Shekhar},
  journal={IEEE Transactions on Computational Social Systems},
  year={2026},
  publisher={IEEE}
}

@inproceedings{tsioutsiouliklis2021fairness,
  title={Fairness-aware pagerank},
  author={Tsioutsiouliklis, Sotiris and Pitoura, Evaggelia and Tsaparas, Panayiotis and Kleftakis, Ilias and Mamoulis, Nikos},
  booktitle={Proceedings of the Web Conference 2021},
  pages={3815--3826},
  year={2021}
}

@article{saxena2025homophily,
  title={Homophily in complex networks: Measures, models, and applications},
  author={Saxena, Akrati and Kumar, Gaurav and Meena, Chandrakala},
  journal={arXiv preprint arXiv:2509.18289},
  year={2025}
}

@book{newman2018networks,
  title={Networks},
  author={Newman, Mark},
  year={2018},
  publisher={Oxford university press}
}

@article{cui2018survey,
  title={A survey on network embedding},
  author={Cui, Peng and Wang, Xiao and Pei, Jian and Zhu, Wenwu},
  journal={IEEE transactions on knowledge and data engineering},
  volume={31},
  number={5},
  pages={833--852},
  year={2018},
  publisher={IEEE}
}

@article{saxena_fairsna_2024,
	title = {{FairSNA}: Algorithmic Fairness in Social Network Analysis},
	volume = {56},
	issn = {0360-0300},
	url = {https://dl.acm.org/doi/10.1145/3653711},
	doi = {10.1145/3653711},
	shorttitle = {{FairSNA}},
	pages = {213:1--213:45},
	number = {8},
	journal = {{ACM} Computing Surveys},
	shortjournal = {{ACM} Comput. Surv.},
	publisher = {{ACM} New York, {NY}},
	author = {Saxena, Akrati and Fletcher, George and Pechenizkiy, Mykola},
	urldate = {2026-05-21},
	year = {2024},
}

@inproceedings{kleindessner_guarantees_2019,
	title = {Guarantees for Spectral Clustering with Fairness Constraints},
	issn = {2640-3498},
	url = {https://proceedings.mlr.press/v97/kleindessner19b.html},
	eventtitle = {International Conference on Machine Learning},
	pages = {3458--3467},
	booktitle = {Proceedings of the 36th International Conference on Machine Learning},
	publisher = {{PMLR}},
	author = {Kleindessner, Matthäus and Samadi, Samira and Awasthi, Pranjal and Morgenstern, Jamie},
	urldate = {2026-06-08},
	year = {2019},
	langid = {english},
}

@inproceedings{palowitch_debiasing_2020,
	title = {Debiasing Graph Representations via Metadata-Orthogonal Training},
	issn = {2473-991X},
	url = {https://ieeexplore.ieee.org/abstract/document/9381348},
	doi = {10.1109/ASONAM49781.2020.9381348},
	eventtitle = {2020 {IEEE}/{ACM} International Conference on Advances in Social Networks Analysis and Mining ({ASONAM})},
	pages = {435--442},
	booktitle = {2020 {IEEE}/{ACM} International Conference on Advances in Social Networks Analysis and Mining ({ASONAM})},
	author = {Palowitch, John and Perozzi, Bryan},
	urldate = {2026-06-08},
	year = {2020},
	note = {{ISSN}: 2473-991X},
}

@misc{li_spectral_2023,
	title = {Spectral Normalized-Cut Graph Partitioning with Fairness Constraints},
	url = {http://arxiv.org/abs/2307.12065},
	doi = {10.3233/FAIA230416},
	author = {Li, Jia and Wang, Yanhao and Merchant, Arpit},
	urldate = {2026-06-08},
	year = {2023},
	eprinttype = {arxiv},
	eprint = {2307.12065 [cs.LG]},
}

@inproceedings{wang_scalable_2023,
	title = {Scalable Spectral Clustering with Group Fairness Constraints},
	issn = {2640-3498},
	url = {https://proceedings.mlr.press/v206/wang23h.html},
	eventtitle = {International Conference on Artificial Intelligence and Statistics},
	pages = {6613--6629},
	booktitle = {Proceedings of The 26th International Conference on Artificial Intelligence and Statistics},
	publisher = {{PMLR}},
	author = {Wang, Ji and Lu, Ding and Davidson, Ian and Bai, Zhaojun},
	urldate = {2026-06-08},
	year = {2023},
	langid = {english},
}

@inproceedings{gupta_consistency_2022,
	title = {Consistency of Constrained Spectral Clustering under Graph Induced Fair Planted Partitions},
	volume = {35},
	url = {https://proceedings.neurips.cc/paper_files/paper/2022/hash/57d7e7e1593ad1ab6818c258fa5654ce-Abstract-Conference.html},
	pages = {13527--13540},
	booktitle = {Advances in Neural Information Processing Systems},
	author = {Gupta, Shubham and Dukkipati, Ambedkar},
	urldate = {2026-06-08},
	year = {2022},
	langid = {english},
}

@article{saxena_nodesim_2022,
	title = {{NodeSim}: node similarity based network embedding for diverse link prediction},
	volume = {11},
	issn = {2193-1127},
	url = {https://epjdatascience.springeropen.com/articles/10.1140/epjds/s13688-022-00336-8},
	doi = {10.1140/epjds/s13688-022-00336-8},
	shorttitle = {{NodeSim}},
	pages = {24},
	number = {1},
	journal = {{EPJ} Data Science},
	shortjournal = {{EPJ} Data Sci.},
	publisher = {Springer},
	author = {Saxena, Akrati and Fletcher, George and Pechenizkiy, Mykola},
	urldate = {2026-06-23},
	year = {2022},
	langid = {english},
}

@inproceedings{buyl_debayes_2020,
	title = {{DeBayes}: a Bayesian Method for Debiasing Network Embeddings},
	issn = {2640-3498},
	url = {https://proceedings.mlr.press/v119/buyl20a.html},
	shorttitle = {{DeBayes}},
	eventtitle = {International Conference on Machine Learning},
	pages = {1220--1229},
	booktitle = {Proceedings of the 37th International Conference on Machine Learning},
	publisher = {{PMLR}},
	author = {Buyl, Maarten and De Bie, Tijl},
	urldate = {2026-06-25},
	year = {2020},
	langid = {english},
}

@inproceedings{he_fairmile_2023,
	location = {New York, {NY}, {USA}},
	title = {{FairMILE}: Towards an Efficient Framework for Fair Graph Representation Learning},
	isbn = {979-8-4007-0381-2},
	url = {https://dl.acm.org/doi/10.1145/3617694.3623231},
	doi = {10.1145/3617694.3623231},
	series = {{EAAMO} '23},
	shorttitle = {{FairMILE}},
	pages = {1--10},
	booktitle = {Proceedings of the 3rd {ACM} Conference on Equity and Access in Algorithms, Mechanisms, and Optimization},
	publisher = {Association for Computing Machinery},
	author = {He, Yuntian and Gurukar, Saket and Parthasarathy, Srinivasan},
	urldate = {2026-06-25},
	year = {2023},
}

@inproceedings{kojaku_residual2vec_2021,
	title = {Residual2Vec: Debiasing graph embedding with random graphs},
	volume = {34},
	url = {https://proceedings.neurips.cc/paper/2021/hash/ca9541826e97c4530b07dda2eba0e013-Abstract.html},
	shorttitle = {Residual2Vec},
	pages = {24150--24163},
	booktitle = {Advances in Neural Information Processing Systems},
	publisher = {Curran Associates, Inc.},
	author = {Kojaku, Sadamori and Yoon, Jisung and Constantino, Isabel and Ahn, Yong-Yeol},
	urldate = {2026-06-25},
	year = {2021},
}

@inproceedings{li_fairlp_2022,
	title = {{FairLP}: Towards Fair Link Prediction on Social Network Graphs},
	volume = {16},
	rights = {Copyright (c) 2022 Association for the Advancement of Artificial Intelligence},
	issn = {2334-0770},
	url = {https://ojs.aaai.org/index.php/ICWSM/article/view/19321},
	doi = {10.1609/icwsm.v16i1.19321},
	shorttitle = {{FairLP}},
	pages = {628--639},
	booktitle = {Proceedings of the International {AAAI} Conference on Web and Social Media},
	author = {Li, Yanying and Wang, Xiuling and Ning, Yue and Wang, Hui},
	urldate = {2026-06-25},
	year = {2022},
	langid = {english},
}

@article{spinelli_fairdrop_2022,
	title = {{FairDrop}: Biased Edge Dropout for Enhancing Fairness in Graph Representation Learning},
	volume = {3},
	issn = {2691-4581},
	url = {https://ieeexplore.ieee.org/abstract/document/9645324},
	doi = {10.1109/TAI.2021.3133818},
	shorttitle = {{FairDrop}},
	pages = {344--354},
	number = {3},
	journal = {{IEEE} Transactions on Artificial Intelligence},
	publisher = {{IEEE}},
	author = {Spinelli, Indro and Scardapane, Simone and Hussain, Amir and Uncini, Aurelio},
	urldate = {2026-06-30},
	year = {2022},
}

@inproceedings{agarwal_towards_2021,
	title = {Towards a unified framework for fair and stable graph representation learning},
	issn = {2640-3498},
	url = {https://proceedings.mlr.press/v161/agarwal21b.html},
	eventtitle = {Uncertainty in Artificial Intelligence},
	pages = {2114--2124},
	booktitle = {Proceedings of the Thirty-Seventh Conference on Uncertainty in Artificial Intelligence},
	publisher = {{PMLR}},
	author = {Agarwal, Chirag and Lakkaraju, Himabindu and Zitnik, Marinka},
	urldate = {2026-07-13},
    year = {2021},
	langid = {english},
}

@inproceedings{cao_fairhelp_2023,
	location = {Cham},
	title = {{FairHELP}: Fairness-Aware Heterogeneous Information Network Embedding for Link Prediction},
	isbn = {978-3-031-30675-4},
	doi = {10.1007/978-3-031-30675-4_22},
	shorttitle = {{FairHELP}},
	pages = {320--330},
	booktitle = {Database Systems for Advanced Applications},
	publisher = {Springer Nature Switzerland},
	author = {Cao, Meng and Song, Jianqing and Yuan, Jinliang and Zhang, Baoming and Wang, Chongjun},
	editor = {Wang, Xin and Sapino, Maria Luisa and Han, Wook-Shin and El Abbadi, Amr and Dobbie, Gill and Feng, Zhiyong and Shao, Yingxiao and Yin, Hongzhi},
	year = {2023},
	langid = {english},
}

@article{cong_fairsample_2024,
	title = {{FairSample}: Training Fair and Accurate Graph Convolutional Neural Networks Efficiently},
	volume = {36},
	issn = {1558-2191},
	url = {https://ieeexplore.ieee.org/abstract/document/10231087},
	doi = {10.1109/TKDE.2023.3306378},
	shorttitle = {{FairSample}},
	pages = {1537--1551},
	number = {4},
	journal = {{IEEE} Transactions on Knowledge and Data Engineering},
	publisher = {{IEEE}},
	author = {Cong, Zicun and Shi, Baoxu and Li, Shan and Yang, Jaewon and He, Qi and Pei, Jian},
	urldate = {2026-07-13},
	year = {2024},
}

@inproceedings{dong_individual_2021,
	location = {New York, {NY}, {USA}},
	title = {Individual Fairness for Graph Neural Networks: A Ranking based Approach},
	isbn = {978-1-4503-8332-5},
	url = {https://dl.acm.org/doi/10.1145/3447548.3467266},
	doi = {10.1145/3447548.3467266},
	series = {{KDD} '21},
	shorttitle = {Individual Fairness for Graph Neural Networks},
	pages = {300--310},
	booktitle = {Proceedings of the 27th {ACM} {SIGKDD} Conference on Knowledge Discovery \& Data Mining},
	publisher = {Association for Computing Machinery},
	author = {Dong, Yushun and Kang, Jian and Tong, Hanghang and Li, Jundong},
	urldate = {2026-07-14},
	year = {2021},
}

@inproceedings{dong_edits_2022,
	location = {New York, {NY}, {USA}},
	title = {{EDITS}: Modeling and Mitigating Data Bias for Graph Neural Networks},
	isbn = {978-1-4503-9096-5},
	url = {https://dl.acm.org/doi/10.1145/3485447.3512173},
	doi = {10.1145/3485447.3512173},
	series = {{WWW} '22},
	shorttitle = {{EDITS}},
	pages = {1259--1269},
	booktitle = {Proceedings of the {ACM} Web Conference 2022},
	publisher = {Association for Computing Machinery},
	author = {Dong, Yushun and Liu, Ninghao and Jalaian, Brian and Li, Jundong},
	urldate = {2026-07-14},
	year = {2022},
}

@inproceedings{kose_fairness-aware_2023,
	title = {Fairness-Aware Graph Filter Design},
	issn = {2576-2303},
	url = {https://ieeexplore.ieee.org/abstract/document/10477040},
	doi = {10.1109/IEEECONF59524.2023.10477040},
	eventtitle = {2023 57th Asilomar Conference on Signals, Systems, and Computers},
	pages = {330--334},
	booktitle = {2023 57th Asilomar Conference on Signals, Systems, and Computers},
	publisher = {{IEEE}},
	author = {Kose, O. Deniz and Shen, Yanning and Mateos, Gonzalo},
	urldate = {2026-07-14},
	year = {2023},
	note = {{ISSN}: 2576-2303},
}

@inproceedings{ling_learning_2022,
	title = {Learning Fair Graph Representations via Automated Data Augmentations},
    booktitle={The Eleventh International Conference on Learning Representations},
	url = {https://openreview.net/forum?id=1_OGWcP1s9w},
	eventtitle = {The Eleventh International Conference on Learning Representations},
	author = {Ling, Hongyi and Jiang, Zhimeng and Luo, Youzhi and Ji, Shuiwang and Zou, Na},
	urldate = {2026-07-14},
	year = {2022},
	langid = {english},
}

@inproceedings{liu_fair_2023,
	location = {New York, {NY}, {USA}},
	title = {Fair Graph Representation Learning via Diverse Mixture-of-Experts},
	isbn = {978-1-4503-9416-1},
	url = {https://dl.acm.org/doi/10.1145/3543507.3583207},
	doi = {10.1145/3543507.3583207},
	series = {{WWW} '23},
	pages = {28--38},
	booktitle = {Proceedings of the {ACM} Web Conference 2023},
	publisher = {Association for Computing Machinery},
	author = {Liu, Zheyuan and Zhang, Chunhui and Tian, Yijun and Zhang, Erchi and Huang, Chao and Ye, Yanfang and Zhang, Chuxu},
	urldate = {2026-07-14},
	year = {2023},
}

@inproceedings{liu_iagnn_2025,
	title = {{IAGNN}: Mitigating Quantity and Topological Imbalance for Fair Graph Learning},
	issn = {2324-9013},
	url = {https://ieeexplore.ieee.org/abstract/document/11354731},
	doi = {10.1109/Trustcom66490.2025.00311},
	shorttitle = {{IAGNN}},
	eventtitle = {2025 {IEEE} 24th International Conference on Trust, Security and Privacy in Computing and Communications ({TrustCom})},
	pages = {2642--2651},
	booktitle = {2025 {IEEE} 24th International Conference on Trust, Security and Privacy in Computing and Communications ({TrustCom})},
	publisher = {{IEEE}},
	author = {Liu, Yangqi and Wang, Xuemin and Chang, Liang and Yang, Heng},
	urldate = {2026-07-14},
	year = {2025},
	note = {{ISSN}: 2324-9013},
}

@inproceedings{bose_compositional_2019,
	title = {Compositional Fairness Constraints for Graph Embeddings},
	issn = {2640-3498},
	url = {https://proceedings.mlr.press/v97/bose19a.html},
	eventtitle = {International Conference on Machine Learning},
	pages = {715--724},
	booktitle = {Proceedings of the 36th International Conference on Machine Learning},
	publisher = {{PMLR}},
	author = {Bose, Avishek and Hamilton, William},
	urldate = {2026-07-14},
	year = {2019},
	langid = {english},
}

@inproceedings{dai_say_2021,
	location = {New York, {NY}, {USA}},
	title = {Say No to the Discrimination: Learning Fair Graph Neural Networks with Limited Sensitive Attribute Information},
	isbn = {978-1-4503-8297-7},
	url = {https://dl.acm.org/doi/10.1145/3437963.3441752},
	doi = {10.1145/3437963.3441752},
	series = {{WSDM} '21},
	shorttitle = {Say No to the Discrimination},
	pages = {680--688},
	booktitle = {Proceedings of the 14th {ACM} International Conference on Web Search and Data Mining},
	publisher = {Association for Computing Machinery},
	author = {Dai, Enyan and Wang, Suhang},
	urldate = {2026-07-14},
	year = {2021},
}

@article{dai_learning_2023,
	title = {Learning Fair Graph Neural Networks With Limited and Private Sensitive Attribute Information},
	volume = {35},
	issn = {1558-2191},
	url = {https://ieeexplore.ieee.org/abstract/document/9864304},
	doi = {10.1109/TKDE.2022.3197554},
	pages = {7103--7117},
	number = {7},
	journal = {{IEEE} Transactions on Knowledge and Data Engineering},
	publisher = {{IEEE}},
	author = {Dai, Enyan and Wang, Suhang},
	urldate = {2026-07-14},
	year = {2023},
}

@misc{guo_fair_2023,
	title = {Fair Attribute Completion on Graph with Missing Attributes},
	url = {http://arxiv.org/abs/2302.12977},
	doi = {10.48550/arXiv.2302.12977},
	number = {{arXiv}:2302.12977},
	publisher = {{arXiv}},
	author = {Guo, Dongliang and Chu, Zhixuan and Li, Sheng},
	urldate = {2026-07-14},
	year = {2023},
	eprinttype = {arxiv},
	eprint = {2302.12977 [cs.LG]},
}

@article{singh_unified_2025,
	title = {A Unified Optimization-Based Framework for Certifiably Robust and Fair Graph Neural Networks},
	volume = {73},
	issn = {1941-0476},
	url = {https://ieeexplore.ieee.org/abstract/document/10789240},
	doi = {10.1109/TSP.2024.3514091},
	pages = {83--98},
	journal = {{IEEE} Transactions on Signal Processing},
	publisher = {{IEEE}},
	author = {Singh, Vipul Kumar and Kumar, Sandeep and Prasad, Avadhesh and {Jayadeva}},
	urldate = {2026-07-14},
	year = {2025},
}

@article{song_mapping_2024,
	title = {{MAPPING}: debiasing graph neural networks for fair node classification with limited sensitive information leakage},
	volume = {27},
	issn = {1573-1413},
	url = {https://doi.org/10.1007/s11280-024-01312-0},
	doi = {10.1007/s11280-024-01312-0},
	shorttitle = {{MAPPING}},
	pages = {74},
	number = {6},
	journal = {World Wide Web},
	shortjournal = {World Wide Web},
	publisher = {Springer},
	author = {Song, Ying and Palanisamy, Balaji},
	urldate = {2026-07-14},
	year = {2024},
	langid = {english},
}

@inproceedings{wang_improving_2022,
	location = {New York, {NY}, {USA}},
	title = {Improving Fairness in Graph Neural Networks via Mitigating Sensitive Attribute Leakage},
	isbn = {978-1-4503-9385-0},
	url = {https://dl.acm.org/doi/10.1145/3534678.3539404},
	doi = {10.1145/3534678.3539404},
	series = {{KDD} '22},
	pages = {1938--1948},
	booktitle = {Proceedings of the 28th {ACM} {SIGKDD} Conference on Knowledge Discovery and Data Mining},
	publisher = {Association for Computing Machinery},
	author = {Wang, Yu and Zhao, Yuying and Dong, Yushun and Chen, Huiyuan and Li, Jundong and Derr, Tyler},
	urldate = {2026-07-14},
	year = {2022},
}

@article{xiao_towards_2025,
	title = {Towards Fair Graph Neural Networks via Counterfactual and Balance},
	volume = {16},
	rights = {http://creativecommons.org/licenses/by/3.0/},
	issn = {2078-2489},
	url = {https://www.mdpi.com/2078-2489/16/8/704},
	doi = {10.3390/info16080704},
	pages = {704},
	number = {8},
	journal = {Information},
	publisher = {Multidisciplinary Digital Publishing Institute},
	author = {Xiao, Zhiguo and Zhou, Yangfan and Li, Dongni and Wang, Ke},
	urldate = {2026-07-14},
	year = {2025},
	langid = {english},
}

@article{zhang_fpgnn_2023,
	title = {{FPGNN}: Fair path graph neural network for mitigating discrimination},
	volume = {26},
	issn = {1573-1413},
	url = {https://doi.org/10.1007/s11280-023-01178-8},
	doi = {10.1007/s11280-023-01178-8},
	shorttitle = {{FPGNN}},
	pages = {3119--3136},
	number = {5},
	journal = {World Wide Web},
	shortjournal = {World Wide Web},
	publisher = {Springer},
	author = {Zhang, Guixian and Cheng, Debo and Zhang, Shichao},
	urldate = {2026-07-14},
	year = {2023},
	langid = {english},
}

@inproceedings{zhang_multi-view_2024,
	location = {Singapore},
	title = {Multi-view Graph Neural Network for Fair Representation Learning},
	isbn = {978-981-97-7238-4},
	doi = {10.1007/978-981-97-7238-4_14},
	pages = {208--223},
	booktitle = {Web and Big Data},
	publisher = {Springer Nature},
	author = {Zhang, Guixian and Yuan, Guan and Cheng, Debo and He, Ludan and Bing, Rui and Li, Jiuyong and Zhang, Shichao},
	editor = {Zhang, Wenjie and Tung, Anthony and Zheng, Zhonglong and Yang, Zhengyi and Wang, Xiaoyang and Guo, Hongjie},
	year = {2024},
	langid = {english},
}

@article{zhang_disentangled_2025,
	title = {Disentangled contrastive learning for fair graph representations},
	volume = {181},
	issn = {0893-6080},
	url = {https://www.sciencedirect.com/science/article/pii/S0893608024007056},
	doi = {10.1016/j.neunet.2024.106781},
	pages = {106781},
	journal = {Neural Networks},
	shortjournal = {Neural Networks},
	publisher = {Elsevier},
	author = {Zhang, Guixian and Yuan, Guan and Cheng, Debo and Liu, Lin and Li, Jiuyong and Zhang, Shichao},
	urldate = {2026-07-14},
	year = {2025},
}

@inproceedings{dam_fairgcl_2026,
	location = {New York, {NY}, {USA}},
	title = {{FairGCL}: Embedding Fairness for Influence Maximization with Graph Contrastive Learning},
	isbn = {979-8-4007-2504-3},
	url = {https://dl.acm.org/doi/10.1145/3795766.3799763},
	doi = {10.1145/3795766.3799763},
	series = {{WebSci} '26},
	shorttitle = {{FairGCL}},
	pages = {227--237},
	booktitle = {Proceedings of the 18th {ACM} Web Science Conference 2026},
	publisher = {Association for Computing Machinery},
	author = {Dam, Arpan and Roy, Sougata and Mitra, Bivas},
	urldate = {2026-07-14},
	year = {2026},
}

@article{hu_migrate_2024,
	title = {Migrate demographic group for fair Graph Neural Networks},
	volume = {175},
	issn = {0893-6080},
	url = {https://www.sciencedirect.com/science/article/pii/S0893608024001886},
	doi = {10.1016/j.neunet.2024.106264},
	pages = {106264},
	journal = {Neural Networks},
	shortjournal = {Neural Networks},
	publisher = {Elsevier},
	author = {Hu, {YanMing} and Liao, {TianChi} and Chen, {JiaLong} and Bian, Jing and Zheng, {ZiBin} and Chen, Chuan},
	urldate = {2026-07-14},
	year = {2024},
}

@article{kose_fair_2022,
	title = {Fair Contrastive Learning on Graphs},
	volume = {8},
	issn = {2373-776X},
	url = {https://ieeexplore.ieee.org/abstract/document/9779533},
	doi = {10.1109/TSIPN.2022.3174953},
	pages = {475--488},
	journal = {{IEEE} Transactions on Signal and Information Processing over Networks},
	publisher = {{IEEE}},
	author = {Kose, Oyku Deniz and Shen, Yanning},
	urldate = {2026-07-14},
	year = {2022},
}

@inproceedings{li_toward_2024,
	location = {New York, {NY}, {USA}},
	title = {Toward Structure Fairness in Dynamic Graph Embedding: A Trend-aware Dual Debiasing Approach},
	isbn = {979-8-4007-0490-1},
	url = {https://dl.acm.org/doi/10.1145/3637528.3671848},
	doi = {10.1145/3637528.3671848},
	series = {{KDD} '24},
	shorttitle = {Toward Structure Fairness in Dynamic Graph Embedding},
	pages = {1701--1712},
	booktitle = {Proceedings of the 30th {ACM} {SIGKDD} Conference on Knowledge Discovery and Data Mining},
	publisher = {Association for Computing Machinery},
	author = {Li, Yicong and Yang, Yu and Cao, Jiannong and Liu, Shuaiqi and Tang, Haoran and Xu, Guandong},
	urldate = {2026-07-14},
	year = {2024},
}

@misc{liu_learning_2025,
	title = {Learning Fair Graph Representations with Multi-view Information Bottleneck},
	url = {http://arxiv.org/abs/2510.25096},
	doi = {10.48550/arXiv.2510.25096},
	number = {{arXiv}:2510.25096},
	publisher = {{arXiv}},
	author = {Liu, Chuxun and Cheng, Debo and Chen, Qingfeng and Gan, Jiangzhang and Li, Jiuyong and Liu, Lin},
	urldate = {2026-07-14},
	year = {2025},
	eprinttype = {arxiv},
	eprint = {2510.25096 [cs.LG]},
}

@article{chen_learning_2026,
	title = {Learning fair graph representation through graph information disentanglement},
	volume = {203},
	issn = {0893-6080},
	url = {https://www.sciencedirect.com/science/article/pii/S0893608026006453},
	doi = {10.1016/j.neunet.2026.109184},
	pages = {109184},
	journal = {Neural Networks},
	shortjournal = {Neural Networks},
	publisher = {Elsevier},
	author = {Chen, Qingfeng and Wei, Wujie and Cheng, Debo and Liu, Chuxun and Jie, Jinyi and Gan, Jiangzhang and Zhang, Shichao},
	urldate = {2026-07-14},
	year = {2026},
}

@inproceedings{guo_towards_2023,
	location = {New York, {NY}, {USA}},
	title = {Towards Fair Graph Neural Networks via Graph Counterfactual},
	isbn = {979-8-4007-0124-5},
	url = {https://dl.acm.org/doi/10.1145/3583780.3615092},
	doi = {10.1145/3583780.3615092},
	series = {{CIKM} '23},
	pages = {669--678},
	booktitle = {Proceedings of the 32nd {ACM} International Conference on Information and Knowledge Management},
	publisher = {Association for Computing Machinery},
	author = {Guo, Zhimeng and Li, Jialiang and Xiao, Teng and Ma, Yao and Wang, Suhang},
	urldate = {2026-07-14},
	year = {2023},
}

@misc{kejani_fair_2024,
	title = {Fair Graph Neural Network with Supervised Contrastive Regularization},
	url = {http://arxiv.org/abs/2404.06090},
	doi = {10.48550/arXiv.2404.06090},
	number = {{arXiv}:2404.06090},
	publisher = {{arXiv}},
	author = {Kejani, Mahdi Tavassoli and Dornaika, Fadi and Loubes, Jean-Michel},
	urldate = {2026-07-14},
	year = {2024},
	eprinttype = {arxiv},
	eprint = {2404.06090 [cs.LG]},
}

@inproceedings{lee_disentangling_2025,
	title = {Disentangling, Amplifying, and Debiasing: Learning Disentangled Representations for Fair Graph Neural Networks},
	volume = {39},
	rights = {Copyright (c) 2025 Association for the Advancement of Artificial Intelligence},
	issn = {2374-3468},
	url = {https://ojs.aaai.org/index.php/AAAI/article/view/33308},
	doi = {10.1609/aaai.v39i11.33308},
	shorttitle = {Disentangling, Amplifying, and Debiasing},
	pages = {12013--12021},
	number = {11},
	booktitle = {Proceedings of the {AAAI} Conference on Artificial Intelligence},
	author = {Lee, Yeon-Chang and Shin, Hojung and Kim, Sang-Wook},
	urldate = {2026-07-14},
	year = {2025},
	langid = {english},
}

@inproceedings{wang_advancing_2024,
	location = {Cham},
	title = {Advancing Graph Counterfactual Fairness Through Fair Representation Learning},
	isbn = {978-3-031-70368-3},
	doi = {10.1007/978-3-031-70368-3_3},
	pages = {40--58},
	booktitle = {Machine Learning and Knowledge Discovery in Databases. Research Track},
	publisher = {Springer Nature Switzerland},
	author = {Wang, Zichong and Chu, Zhibo and Blanco, Ronald and Chen, Zhong and Chen, Shu-Ching and Zhang, Wenbin},
	editor = {Bifet, Albert and Davis, Jesse and Krilavičius, Tomas and Kull, Meelis and Ntoutsi, Eirini and Žliobaitė, Indrė},
	year = {2024},
	langid = {english},
}

@inproceedings{wang_fair_2026,
	title = {Fair Graph Learning with Limited Sensitive Attribute Information},
	volume = {40},
	rights = {Copyright (c) 2026 Association for the Advancement of Artificial Intelligence},
	issn = {2374-3468},
	url = {https://ojs.aaai.org/index.php/AAAI/article/view/41292},
	doi = {10.1609/aaai.v40i46.41292},
	pages = {39423--39431},
	number = {46},
	booktitle = {Proceedings of the {AAAI} Conference on Artificial Intelligence},
	author = {Wang, Zichong and Yang, Jie and Zhuang, Jun and Jiang, Puqing and Chen, Mingzhe and Hu, Ye and Zhang, Wenbin},
	urldate = {2026-07-14},
	year = {2026},
	langid = {english},
}

@inproceedings{zhu_fair_2024,
	location = {New York, {NY}, {USA}},
	title = {Fair Graph Representation Learning via Sensitive Attribute Disentanglement},
	isbn = {979-8-4007-0171-9},
	url = {https://dl.acm.org/doi/10.1145/3589334.3645532},
	doi = {10.1145/3589334.3645532},
	series = {{WWW} '24},
	pages = {1182--1192},
	booktitle = {Proceedings of the {ACM} Web Conference 2024},
	publisher = {Association for Computing Machinery},
	author = {Zhu, Yuchang and Li, Jintang and Zheng, Zibin and Chen, Liang},
	urldate = {2026-07-14},
	year = {2024},
}

@inproceedings{jiang_chasing_2024,
	title = {Chasing Fairness in Graphs: A {GNN} Architecture Perspective},
	volume = {38},
	rights = {Copyright (c) 2024 Association for the Advancement of Artificial Intelligence},
	issn = {2374-3468},
	url = {https://ojs.aaai.org/index.php/AAAI/article/view/30115},
	doi = {10.1609/aaai.v38i19.30115},
	shorttitle = {Chasing Fairness in Graphs},
	pages = {21214--21222},
	number = {19},
	booktitle = {Proceedings of the {AAAI} Conference on Artificial Intelligence},
	author = {Jiang, Zhimeng and Han, Xiaotian and Fan, Chao and Liu, Zirui and Zou, Na and Mostafavi, Ali and Hu, Xia},
	urldate = {2026-07-14},
	year = {2024},
	langid = {english},
}

@article{kose_fairgat_2024,
	title = {{FairGAT}: Fairness-Aware Graph Attention Networks},
	volume = {18},
	issn = {1556-4681},
	url = {https://dl.acm.org/doi/10.1145/3645096},
	doi = {10.1145/3645096},
	shorttitle = {{FairGAT}},
	pages = {164:1--164:20},
	number = {7},
	journal = {{ACM} Transactions on Knowledge Discovery from Data},
	shortjournal = {{ACM} Trans. Knowl. Discov. Data},
	publisher = {{ACM} New York, {NY}},
	author = {Kose, O. Deniz and Shen, Yanning},
	urldate = {2026-07-14},
	year = {2024},
}

@inproceedings{lin_bemap_2024,
	title = {{BeMap}: Balanced Message Passing for Fair Graph Neural Network},
	issn = {2640-3498},
	url = {https://proceedings.mlr.press/v231/lin24a.html},
	shorttitle = {{BeMap}},
	eventtitle = {Learning on Graphs Conference},
	pages = {37:1--37:25},
	booktitle = {Proceedings of the Second Learning on Graphs Conference},
	publisher = {{PMLR}},
	author = {Lin, Xiao and Kang, Jian and Cong, Weilin and Tong, Hanghang},
	urldate = {2026-07-14},
	year = {2024},
	langid = {english},
}

@article{liu_generalized_2023,
	title = {On Generalized Degree Fairness in Graph Neural Networks},
	volume = {37},
	rights = {Copyright (c) 2023 Association for the Advancement of Artificial Intelligence},
	issn = {2374-3468},
	url = {https://ojs.aaai.org/index.php/AAAI/article/view/25574},
	doi = {10.1609/aaai.v37i4.25574},
	pages = {4525--4533},
	number = {4},
	journal = {Proceedings of the {AAAI} Conference on Artificial Intelligence},
	author = {Liu, Zemin and Nguyen, Trung-Kien and Fang, Yuan},
	urldate = {2026-07-14},
	year = {2023},
	langid = {english},
}

@inproceedings{luo_fairgt_2024,
	title = {{FairGT}: A Fairness-aware Graph Transformer},
	isbn = {978-1-956792-04-1},
	url = {https://dl.acm.org/doi/abs/10.24963/ijcai.2024/50},
	doi = {10.24963/ijcai.2024/50},
	series = {Guide Proceedings},
	pages = {449--457},
	booktitle = {Proceedings of the Thirty-Third International Joint Conference on Artificial Intelligence},
	author = {Luo, Renqiang and Huang, Huafei and Yu, Shuo and Zhang, Xiuzhen and Xia, Feng},
	urldate = {2026-07-14},
	year = {2024},
}

@inproceedings{purificato_gnns_2025,
	location = {New York, {NY}, {USA}},
	title = {{GNN}’s {FAME}: Fairness-Aware {MEssages} for Graph Neural Networks},
	isbn = {979-8-4007-1313-2},
	url = {https://dl.acm.org/doi/10.1145/3699682.3728324},
	doi = {10.1145/3699682.3728324},
	series = {{UMAP} '25},
	shorttitle = {{GNN}’s {FAME}},
	pages = {301--306},
	booktitle = {Proceedings of the 33rd {ACM} Conference on User Modeling, Adaptation and Personalization},
	publisher = {Association for Computing Machinery},
	author = {Purificato, Erasmo and Mahadik, Hannan Javed and Boratto, Ludovico and De Luca, Ernesto William},
	urldate = {2026-07-14},
	year = {2025},
}

@article{wang_toward_2024,
	title = {Toward fair graph neural networks via real counterfactual samples},
	volume = {66},
	issn = {0219-3116},
	url = {https://doi.org/10.1007/s10115-024-02161-z},
	doi = {10.1007/s10115-024-02161-z},
	pages = {6617--6641},
	number = {11},
	journal = {Knowledge and Information Systems},
	shortjournal = {Knowl Inf Syst},
	publisher = {Springer},
	author = {Wang, Zichong and Qiu, Meikang and Chen, Min and Ben Salem, Malek and Yao, Xin and Zhang, Wenbin},
	urldate = {2026-07-14},
	year = {2024},
	langid = {english},
}

@article{wang_fair_2025,
	title = {Fair Graph U-Net: A Fair Graph Learning Framework Integrating Group and Individual Awareness},
	volume = {39},
	rights = {Copyright (c) 2025 Association for the Advancement of Artificial Intelligence},
	issn = {2374-3468},
	url = {https://ojs.aaai.org/index.php/AAAI/article/view/35071},
	doi = {10.1609/aaai.v39i27.35071},
	shorttitle = {Fair Graph U-Net},
	pages = {28485--28493},
	number = {27},
	journal = {Proceedings of the {AAAI} Conference on Artificial Intelligence},
	author = {Wang, Zichong and Chu, Zhibo and Doan, Thang Viet and Wang, Shaowei and Wu, Yongkai and Palade, Vasile and Zhang, Wenbin},
	urldate = {2026-07-14},
	year = {2025},
	langid = {english},
}

@misc{zhu_fairness-aware_2023,
	title = {Fairness-aware Message Passing for Graph Neural Networks},
	url = {http://arxiv.org/abs/2306.11132},
	doi = {10.48550/arXiv.2306.11132},
	number = {{arXiv}:2306.11132},
	publisher = {{arXiv}},
	author = {Zhu, Huaisheng and Fu, Guoji and Guo, Zhimeng and Zhang, Zhiwei and Xiao, Teng and Wang, Suhang},
	urldate = {2026-07-14},
	year = {2023},
	eprinttype = {arxiv},
	eprint = {2306.11132 [cs.LG]},
}

@article{zhu_fairagg_2024,
	title = {{FairAGG}: Toward Fair Graph Neural Networks via Fair Aggregation},
	volume = {11},
	issn = {2329-924X},
	url = {https://ieeexplore.ieee.org/abstract/document/10516580},
	doi = {10.1109/TCSS.2024.3385539},
	shorttitle = {{FairAGG}},
	pages = {6308--6319},
	number = {5},
	journal = {{IEEE} Transactions on Computational Social Systems},
	publisher = {{IEEE}},
	author = {Zhu, Yuchang and Li, Jintang and Chen, Liang and Zheng, Zibin},
	urldate = {2026-07-14},
	year = {2024},
}

@inproceedings{wang_fairgnn-wod_2025,
	title = {{fairGNN}-{WOD}: fair graph learning without demographics},
	isbn = {978-1-956792-06-5},
	url = {https://dl.acm.org/doi/abs/10.24963/ijcai.2025/63},
	doi = {10.24963/ijcai.2025/63},
	series = {Guide Proceedings},
	pages = {556--564},
	booktitle = {Proceedings of the Thirty-Fourth International Joint Conference on Artificial Intelligence},
	author = {Wang, Zichong and Liu, Fang and Pan, Shimei and Liu, Jun and Saeed, Fahad and Qiu, Meikang and Zhang, Wenbin},
	urldate = {2026-07-14},
	year = {2025},
}

@inproceedings{wang_towards_2025-1,
	title = {Towards Fair Graph Neural Networks via Graph Counterfactual Without Sensitive Attributes},
	issn = {2375-026X},
	url = {https://ieeexplore.ieee.org/abstract/document/11112997},
	doi = {10.1109/ICDE65448.2025.00027},
	eventtitle = {2025 {IEEE} 41st International Conference on Data Engineering ({ICDE})},
	pages = {265--277},
	booktitle = {2025 {IEEE} 41st International Conference on Data Engineering ({ICDE})},
	publisher = {{IEEE}},
	author = {Wang, Xuemin and Gu, Tianlong and Bao, Xuguang and Chang, Liang},
	urldate = {2026-07-14},
	year = {2025},
	note = {{ISSN}: 2375-026X},
}

@inproceedings{zhu_devil_2024,
	location = {New York, {NY}, {USA}},
	title = {The Devil is in the Data: Learning Fair Graph Neural Networks via Partial Knowledge Distillation},
	isbn = {979-8-4007-0371-3},
	url = {https://dl.acm.org/doi/10.1145/3616855.3635768},
	doi = {10.1145/3616855.3635768},
	series = {{WSDM} '24},
	shorttitle = {The Devil is in the Data},
	pages = {1012--1021},
	booktitle = {Proceedings of the 17th {ACM} International Conference on Web Search and Data Mining},
	publisher = {Association for Computing Machinery},
	author = {Zhu, Yuchang and Li, Jintang and Chen, Liang and Zheng, Zibin},
	urldate = {2026-07-14},
	year = {2024},
}

@article{li_toward_2026,
	title = {Toward fair graph neural networks via dual-teacher knowledge distillation},
	volume = {194},
	issn = {0893-6080},
	url = {https://www.sciencedirect.com/science/article/pii/S0893608025010640},
	doi = {10.1016/j.neunet.2025.108184},
	pages = {108184},
	journal = {Neural Networks},
	shortjournal = {Neural Networks},
	publisher = {Elsevier},
	author = {Li, Chengyu and Cheng, Debo and Zhang, Guixian and Li, Yi and Zhang, Shichao},
	urldate = {2026-07-14},
	year = {2026},
}

@article{wang_generating_2024,
	title = {Generating Diagnostic and Actionable Explanations for Fair Graph Neural Networks},
	volume = {38},
	rights = {Copyright (c) 2024 Association for the Advancement of Artificial Intelligence},
	issn = {2374-3468},
	url = {https://ojs.aaai.org/index.php/AAAI/article/view/30168},
	doi = {10.1609/aaai.v38i19.30168},
	pages = {21690--21698},
	number = {19},
	journal = {Proceedings of the {AAAI} Conference on Artificial Intelligence},
	author = {Wang, Zhenzhong and Zeng, Qingyuan and Lin, Wanyu and Jiang, Min and Tan, Kay Chen},
	urldate = {2026-07-14},
	year = {2024},
	langid = {english},
}

@article{zhang_learning_2024,
	title = {Learning fair representations via rebalancing graph structure},
	volume = {61},
	issn = {0306-4573},
	url = {https://www.sciencedirect.com/science/article/pii/S0306457323003072},
	doi = {10.1016/j.ipm.2023.103570},
	pages = {103570},
	number = {1},
	journal = {Information Processing \& Management},
	shortjournal = {Information Processing \& Management},
	publisher = {Elsevier},
	author = {Zhang, Guixian and Cheng, Debo and Yuan, Guan and Zhang, Shichao},
	urldate = {2026-07-14},
	year = {2024},
}

@inproceedings{zhu_one_2024,
	location = {New York, {NY}, {USA}},
	title = {One Fits All: Learning Fair Graph Neural Networks for Various Sensitive Attributes},
	isbn = {979-8-4007-0490-1},
	url = {https://dl.acm.org/doi/10.1145/3637528.3672029},
	doi = {10.1145/3637528.3672029},
	series = {{KDD} '24},
	shorttitle = {One Fits All},
	pages = {4688--4699},
	booktitle = {Proceedings of the 30th {ACM} {SIGKDD} Conference on Knowledge Discovery and Data Mining},
	publisher = {Association for Computing Machinery},
	author = {Zhu, Yuchang and Li, Jintang and Bian, Yatao and Zheng, Zibin and Chen, Liang},
	urldate = {2026-07-14},
	year = {2024},
}

@inproceedings{cao_flexible_2023,
	title = {A Flexible Debiasing Framework for Fair Heterogeneous Information Network Embedding},
	url = {https://journals.sagepub.com/action/showAbstract},
	doi = {10.3233/FAIA230289},
	booktitle = {{ECAI} 2023},
	publisher = {{SAGE} Publications},
	author = {Cao, Meng and Chen, Mingcai and Song, Jianqing and Fang, Chenxuan and Wang, Chongjun},
	editor = {Gal, Kobi and Nowé, Ann and Nalepa, Grzegorz J. and Fairstein, Roy and Rădulescu, Roxana},
	urldate = {2026-07-14},
	year = {2023},
}

@misc{cao_fahin_2024,
	location = {Rochester, {NY}},
	title = {Fahin: A Unified Framework for Fair Representation Learning on Heterogeneous Information Networks},
	url = {https://papers.ssrn.com/abstract=4918430},
	shorttitle = {Fahin},
	number = {4918430},
	publisher = {Social Science Research Network},
	author = {Cao, Meng and Yu, Hualei and Chen, Mingcai and Song, Jianqing and Wang, Chongjun},
	urldate = {2026-07-14},
	year = {2024},
	langid = {english},
}

@inproceedings{fan_fair_2021,
	title = {Fair Graph Auto-Encoder for Unbiased Graph Representations with Wasserstein Distance},
	issn = {2374-8486},
	url = {https://ieeexplore.ieee.org/abstract/document/9679109},
	doi = {10.1109/ICDM51629.2021.00122},
	eventtitle = {2021 {IEEE} International Conference on Data Mining ({ICDM})},
	pages = {1054--1059},
	booktitle = {2021 {IEEE} International Conference on Data Mining ({ICDM})},
	publisher = {{IEEE}},
	author = {Fan, Wei and Liu, Kunpeng and Xie, Rui and Liu, Hao and Xiong, Hui and Fu, Yanjie},
	urldate = {2026-07-14},
	year = {2021},
	note = {{ISSN}: 2374-8486},
}

@misc{kose_fairnorm_2022,
	title = {{FairNorm}: Fair and Fast Graph Neural Network Training},
	url = {http://arxiv.org/abs/2205.09977},
	doi = {10.48550/arXiv.2205.09977},
	shorttitle = {{FairNorm}},
	number = {{arXiv}:2205.09977},
	publisher = {{arXiv}},
	author = {Kose, O. Deniz and Shen, Yanning},
	urldate = {2026-07-14},
	year = {2022},
	eprinttype = {arxiv},
	eprint = {2205.09977 [cs.LG]},
}

@inproceedings{ma_learning_2022,
	location = {New York, {NY}, {USA}},
	title = {Learning Fair Node Representations with Graph Counterfactual Fairness},
	isbn = {978-1-4503-9132-0},
	url = {https://dl.acm.org/doi/10.1145/3488560.3498391},
	doi = {10.1145/3488560.3498391},
	series = {{WSDM} '22},
	pages = {695--703},
	booktitle = {Proceedings of the Fifteenth {ACM} International Conference on Web Search and Data Mining},
	publisher = {Association for Computing Machinery},
	author = {Ma, Jing and Guo, Ruocheng and Wan, Mengting and Yang, Longqi and Zhang, Aidong and Li, Jundong},
	urldate = {2026-07-14},
	year = {2022},
}

@misc{navarin_learning_2020,
	title = {Learning deep fair graph neural networks},
	url = {https://www.amazon.science/publications/learning-deep-fair-graph-neural-networks},
	titleaddon = {Amazon Science},
	author = {Navarin, Nicolo and Oneto, Luca and Donini, Michele},
	urldate = {2026-07-14},
	year = {2020},
	langid = {english},
}

@inproceedings{luo_fairge_2026,
	location = {New York, {NY}, {USA}},
	title = {{FairGE}: Fairness-Aware Graph Encoding in Incomplete Social Networks},
	isbn = {979-8-4007-2307-0},
	url = {https://dl.acm.org/doi/10.1145/3774904.3792169},
	doi = {10.1145/3774904.3792169},
	series = {{WWW} '26},
	shorttitle = {{FairGE}},
	pages = {4541--4552},
	booktitle = {Proceedings of the {ACM} Web Conference 2026},
	publisher = {Association for Computing Machinery},
	author = {Luo, Renqiang and Huang, Huafei and Tang, Tao and Ren, Jing and Xu, Ziqi and Hou, Mingliang and Dai, Enyan and Xia, Feng},
	urldate = {2026-07-14},
	year = {2026},
}

@inproceedings{wang_fairness-aware_2026,
	location = {Cham},
	title = {Fairness-Aware Graph Representation Learning with Limited Demographic Information},
	isbn = {978-3-032-05962-8},
	doi = {10.1007/978-3-032-05962-8_21},
	pages = {354--371},
	booktitle = {Machine Learning and Knowledge Discovery in Databases. Research Track},
	publisher = {Springer Nature Switzerland},
	author = {Wang, Zichong and Yin, Zhipeng and Yang, Liping and Zhuang, Jun and Yu, Rui and Kong, Qingzhao and Zhang, Wenbin},
	editor = {Ribeiro, Rita P. and Pfahringer, Bernhard and Japkowicz, Nathalie and Larrañaga, Pedro and Jorge, Alípio M. and Soares, Carlos and Abreu, Pedro H. and Gama, João},
	year = {2026},
	langid = {english},
}

@article{gholinejad_heterophily-aware_2026,
	title = {Heterophily-aware fair recommendation using graph convolutional networks},
	volume = {661},
	issn = {0925-2312},
	url = {https://www.sciencedirect.com/science/article/pii/S0925231225026281},
	doi = {10.1016/j.neucom.2025.131956},
	pages = {131956},
	journal = {Neurocomputing},
	shortjournal = {Neurocomputing},
	publisher = {Elsevier},
	author = {Gholinejad, Nemat and Chehreghani, Mostafa Haghir},
	urldate = {2026-07-14},
	year = {2026},
}

\end{document}